\documentclass{elsart3p}
\def \irbaddress{Rudjer Bo\v{s}kovi\'{c} Institute,
Bijeni\v{c}ka cesta 54, P.O. Box 180, 10002 Zagreb, Croatia}
\def \ksuaddress{Department of Physics, Kent State University, 105 Smith Hall, Kent, OH 44242, USA}
\def \fsuaddress{Department of Physics, Florida State University, 
Tallahassee, FL 32306-4350, USA}

\usepackage[usenames]{color}
\usepackage{epsfig}
\usepackage{amssymb,amsmath}

\usepackage[usenames]{color}

\usepackage{amssymb}

\begin{document}

\begin{frontmatter}



\title{Model-independent resonance parameter extraction \\using the trace of K and T matrices}


\author[IRB]{S. Ceci},
\author[IRB]{A. \v Svarc\corauthref{ca}},
\corauth[ca]{Corresponding author}
\ead{svarc@irb.hr}
\author[IRB]{B. Zauner},
\author[KSU]{D. M. Manley} \and
\author[FSU]{S. Capstick}
\address[IRB]{\irbaddress}
\address[KSU]{\ksuaddress}
\address[FSU]{\fsuaddress}

\begin{abstract}
A model-independent method for the determination of Breit-Wigner
resonance parameters is presented. The method is based on eliminating
the dependence on the choice of channel basis by analyzing the trace
of the $K$ and $T$ matrices in the coupled-channel formalism, rather
than individual matrix elements of the multichannel scattering
matrix. 
\end{abstract}

\end{frontmatter}

\section{Introduction}
A general problem in theoretical baryon physics is to make a
connection between resonances that are predicted by various quark
models and experiment. A reasonable way to proceed is by identifying
the poles of analytic functions that are able to describe
simultaneously all experimental data in a multiplicity of existing
channels with theoretically predicted resonant states. Therefore,
properly and uniquely extracting resonance parameters from experiment
is a task of primary importance. We emphasize the problem of
uniqueness.  The work described here is motivated by the need to
extract Breit-Wigner resonance parameters from multichannel
partial-wave analyses (PWAs) in a model-independent way. Many PWAs of
similar experimental data produce similar partial waves, while the
extracted Breit-Wigner resonance parameters are often quite
different. This fact can easily be seen in the {\it Review of Particle
Physics}~\cite{PDG04} by the Particle Data Group (PDG). Each resonance
in the {\it Review} has been parametrized in two ways. First, there
are Breit-Wigner parameters, i.e. the resonance mass $M^R$, decay
width $\Gamma^R$, and branching ratios $x_a$ (the ratio between the
partial width into channel $a$ and the total width). Alternatively,
$T$-matrix complex poles ($\mathrm {Re}\, W_p$ and $-2\,\mathrm {Im}\,
W_p$), as well as complex residues (moduli $|r|$, and phases $\theta$)
are given in~\cite{PDG04}. Breit-Wigner parameters obtained in various
partial-wave analyses vary quite substantially, partly because details
of those analyses are different; i.e. the number and character of the
included channels, different parameterization schemes, analyticity
constraints for scattering amplitudes, the choice of background
models, and the method of unitarization (if at all) of the $S$
matrix. However, it is also the case that the methods for extracting
resonance parameters are different: Argand-plot fits~\cite{Lon74},
Breit-Wigner fits with background~\cite{Arn04}, direct fits of
analysis parameters~\cite{Man91,Man95}, or model specific schemes
which extract $T$ matrix poles~\cite{Cut79,Bat95,Vra00}.

In this Letter we present a model-independent method for extracting
Breit-Wigner parameters from any unitary multichannel analysis able to
provide the full $T$ matrix, using the trace of the corresponding $K$
matrix. Since all Breit-Wigner parameterizations are equivalent at the
energy of the $K$-matrix pole, the parameters obtained using this
method should be directly compared to those from quark models and
lattice QCD. In order to connect the results of a model-independent
$K$ matrix extraction with those of a model-dependent analysis,
e.g. based on the $T$ matrix, we shall keep the relations defining
multichannel $T$ and $K$ matrices as general as possible. It turns out
that the $T$-matrix trace simplifies the formalism without loss of
generality, and shows resonant behavior more prominently than any $T$
matrix element does. To illustrate this, we shall take the $T$ matrix
from an earlier analysis~\cite{Bat95} and recalculate the resonance
parameters. The $T$-matrix trace happens to show resonant behavior at
energies matching those of the $K$-matrix poles.

\section{ Multichannel scattering} 
The essence of any multichannel theory is the fact that the evolution
of a system is no longer described by scalars, but by operators acting
in an orthonormal wave-function space, and the transition
probabilities for physical (measurable) processes are given by the
matrix elements of their representation in the chosen basis. Once this
basis is specified, the evolution of the system is described by
solving equations which are matrices in the multichannel space,
rather than scalar equations.

All equations given here are considered to be matrix relations, unless
matrix indices are explicitly stated. The transition probability
$P_{a\rightarrow b}$ that a two-body system from initial channel $|a;q
\rangle $ ends up in the final two-body (or quasi-two-body) channel
$|b;q\rangle$ is given by the absolute square of the scattering
$S^{q}$-matrix element \nolinebreak[4]{$P_{a\rightarrow b}=|\langle
b;q|\hat{S}^{q}|a;q\rangle|^2$}, where $q$ designates all quantum
numbers conserved in the scattering reaction, and $a$ and $b$ are
channels. In the case of $\pi N$ scattering we have conserved spin,
parity, and almost conserved isospin (charge symmetry is only slightly
violated). Conservation of probability is ensured if the $S$ matrix
(for simplicity, we drop $q$ henceforth) is unitary. Therefore, the $S$
matrix can be written as {\nolinebreak $S=e^{2i\delta}$}, where
$\delta$ is some matrix Hermitian in the channel indices. Hermitian
matrices have real eigenvalues and are diagonalized by unitary
matrices. The $\delta$ matrix is related to a real, diagonal matrix
$\delta_D$ by a unitary transformation {\nolinebreak $\delta=U^\dag
\delta_D U$}, where $U$ is a unitary matrix. The $S$ matrix is
evidently diagonalized by the same transformation, so {\nolinebreak
$S=U^\dag e^{2i\delta_D} U$}.

The $K$ matrix~\cite{Cut79,MarSpe} is defined as {\nolinebreak
$K=i(I-S)/(I+S)$}, where $I$ is the unit matrix. The $K$ matrix can,
in the eigenstate basis, be written using the diagonal matrix
$\delta_D$ as \nolinebreak{ $K=U^\dag \tan \delta_D U$}. The $K$
matrix is Hermitian because $S$ is unitary, and symmetric because of
time-reversal invariance, so $K$ is, in fact, a real matrix. Thus, $U$
is a real orthogonal matrix that we henceforth designate as $O$.

Every diagonal $N\times N$ matrix can be spanned in the ortho-normal
vector basis $\{E^1,\ldots,E^N\}$:
\begin{equation}
E^1=\begin{pmatrix}
1 & 0 & \cdot & 0 \\ 
0 & 0 & \cdot & \cdot \\ 
\vdots & \vdots & \ddots & \vdots\\
 0 & \cdot& \cdot & 0
\end{pmatrix},\ \cdots \ ,\
E^N=\begin{pmatrix}
0 & \cdot & \cdot & 0 \\ 
\vdots & \ddots & \vdots & \vdots \\
\cdot & \cdot & 0 & 0 \\ 
0 & \cdot& 0 & 1
\end{pmatrix},
\end{equation}
so, in our case, we have
\begin{equation}
\tan \delta_D = \sum_{i=1}^{N}E^i\,\tan\delta^i ,
\end{equation}
where $\delta^i$ is the $i$th diagonal element of $ \delta_D$, also
known as the {\it eigenphase shift}, and $N$ is the number of
channels. We define the coupling matrices $\chi^i$ to be
\begin{equation}\label{eq:chiDefinition}
\chi^i = O^T E^i O,
\end{equation}
and these matrices turn out to be ortho-normal projectors:
\begin{equation}
\sum_{i=1}^N \chi^i = I, \ \
\chi^i\ \chi^j\ = \chi^j \delta_{ij},
\end{equation}
where $\delta_{ij}$ is the Kronecker $\delta$ symbol. 

The trace of a matrix is, by definition, a sum of its diagonal
elements. A trace has two particularly important properties: i) the
trace of a product of matrices is invariant with respect to cyclic
permutations, $\mathrm{Tr}\,[ABC]\,=\,\mathrm{Tr}\,[BCA]$; and ii) the trace
is a distributive function with respect to scalars $\alpha$ and
$\beta$, $\mathrm{Tr}\,[\alpha\, A+\beta\,B]=\alpha\,\mathrm{Tr}\,[A]+\beta
\,\mathrm{Tr}\,[B]$.

The orthogonal transformation in definition~(\ref{eq:chiDefinition})
conserves the trace of a matrix, so 
\begin{equation}\label{eq:TraceIsOne}
\mathrm{Tr}\, \chi^i\, =\, 1.
\end{equation}
It follows that the matrices $K$ and $T$ can be written as the sums
\begin{equation}
K=\sum_{j=1}^{N} \chi^j \ \tan\delta^j, \ \ \
T=\sum_{j=1}^{N} \chi^j \ e^{i\delta^j}\sin\delta^j,
\end{equation} 
where the connection between the $K$ and $T$ matrices is given by the
relation
\begin{equation}\label{eq:defToverK}
K=T/(I+iT).
\end{equation}

\section{Breit-Wigner parameterization} 
Elements of $\tan \delta_D$, as well as the $\chi^j$, are functions of
energy or a corresponding kinematical variable, and their description
requires modeling of the energy dependence of numerous functions. We
see resonances in scattering reactions as real poles of the $K$
matrix. The $r$th element of the diagonal matrix $\tan \delta_D$
can be written as~\cite{MarSpe}
\begin{equation}\label{eq:tanBW}
\tan \delta^r = \frac{\Gamma_r/2}{M_r-W}+\tan \delta^r_B,
\end{equation}
where the selected pole term is parametrized in Breit-Wigner form, and
it is singled out from other contributions, designated collectively as
the background term at resonance $\tan \delta^r_B$. The Breit-Wigner
mass ($M_r$) and total width ($\Gamma_r$) parameters are allowed to be
functions of the center-of-mass total energy $W$. The reported Breit-Wigner
parameters $M^R_r$ and $\Gamma^R_r$ are given by the values of
$M_r(W)$ and $\Gamma_r(W)$ evaluated at an
energy equal to the corresponding resonance mass $M^R_r$:
\begin{equation}
M^R_r=M_r\!\left(M^R_r\right),\ \ \
\Gamma^R_r=\Gamma_r\!\left(M^R_r\right),
\end{equation}
where we have explicitly written $M_r$ and $\Gamma_r$ from
Eq.~(\ref{eq:tanBW}) as functions of energy $W$.

The corresponding $K$  and $T$ matrices are given by the equations
\begin{eqnarray}\label{eq:KMartin}
K &=& \chi^r \frac{\Gamma_r'/2}{M_r-W} + \sum_{j\neq r}^{N} \chi^j \tan \delta^j,\\
\label{eq:TMartin}T &=& \chi^r \frac{\Gamma_r'/2}{M_r-W-i\Gamma_r'/2} + \sum_{j\neq r}^{N} \chi^j e^{i\delta^j} 
\sin \delta^j,
\end{eqnarray} 
where the second term in each equation is the coupled-channel
background contribution, and $\Gamma_r'/2$ represents
$\Gamma_r/2+(M_r-W) \tan \delta_B^r$. When $W$ equals the mass of the
resonance, $\Gamma_r'$ is manifestly equal to $\Gamma_r$. Although
these relations are in general a sum over several resonances $r$, here
they are written for one resonance for simplicity.

If there is a pole in the $K$ matrix at some energy $M^R_r$, then the
matrix element $\chi^r_{ab}$ at that energy gives the coupling
strength of the resonance with mass $M^R_r$ and total decay width
$\Gamma^R_r$ from channel $a$ to channel $b$. The diagonal element of
the matrix $\chi^r$ is the branching ratio $x^r_{a}$ of a given
resonance to the channel $a$
\begin{equation}\label{eq:defBranchingRatio}
x^r_a=\chi^r_{aa}.
\end{equation}

\section{Extraction procedure} 
The channel dependence of resonance parameters can be reduced
significantly by using only diagonal elements of the $T$ and $K$
matrices. In practice, these matrices can be obtained either by
unitary coupled-channel partial-wave analyses, or by using
partial-wave $T$ matrices obtained in diverse single-channel PWAs as
input to a unitary coupled-channel formalism, and refitting them to
obtain a unitary set of all coupled-channel $T$ matrix elements.

Channel dependence is completely removed from the sums
\begin{equation}
\mathrm{Tr}(K)= \sum_{j=1}^{N} \tan \delta^j, \ \ \ \mathrm{Tr}(T) = \sum_{j=1}^{N} e^{i\delta^j}\sin \delta^j,
\end{equation}
because the traces of the $K$ and $T$ matrices are the same as the
traces of their similar diagonal partners $\tan \delta_D$ and
$e^{i\delta_D}\sin \delta_D$, respectively. The same is also evident
from Eq.~(\ref{eq:TraceIsOne}). Consequently, Eqs.~(\ref{eq:KMartin})
and~(\ref{eq:TMartin}) are simplified by taking the traces
\begin{eqnarray}
\label{eq:Ktrace}\mathrm{Tr}(K) &=&\frac{\Gamma_r'/2}{M_r-W} + \sum_{j\neq r}^{N} \tan \delta^j,\\
\mathrm{Tr}(T) &=& \frac{\Gamma_r'/2}{M_r-W-i\Gamma_r'/2} + \sum_{j\neq r}^{N} e^{i\delta^j}\sin \delta^j.
\end{eqnarray}
The last relation, i.e.\ the $T$-matrix trace, would be a good
starting point for model-dependent extraction methods. However,
instead of putting considerable effort into modeling the background
and energy- and channel-dependent resonance parameters, we use the
following procedure:
\begin{enumerate}
\item The parameter extraction procedure starts when a full $T$ matrix has been obtained 
from an energy-dependent partial-wave analysis of experimental data.
\item Contrary to the usual prescription, where
  Eq.~({\ref{eq:TMartin}}) is used to obtain resonance parameters from
  the $T$ matrix in a model-dependent way, we use
  Eq.~(\ref{eq:defToverK}) to obtain the full $K$ matrix from the
  known $T$ matrix.
\item Poles of $\mathrm{Tr}\, K$ are found to obtain a set of resonance
  masses $M^R_1,\cdots M^R_{N_R}$, where $N_R$ is the number of
  resonances.
\item Multiplying both sides of Eq.~(\ref{eq:Ktrace}) by
  $(M^R_k-\nolinebreak W)$ and setting the energy $W$ to the value of
  the $k$th resonance mass $(M^R_k)$, the corresponding resonance
  width is isolated:
\begin{equation}
\Gamma^R_k=2 \lim_{W\rightarrow M^R_k}\bigg[ \left(M^R_k-W\right)\, \mathrm{Tr}(K)\bigg].
\end{equation}
  All other contributions to the $K$ matrix trace, i.e. background,
  other resonances, and channel-couplings, are removed in this limiting
  process (this relation turns out to be similar to Eq.(16) in
  Ref.~\cite{Gri04} for the case of the various $\pi N$ isospin
  channels).
\item The branching ratio of a resonance to a given channel can be
  obtained in similar manner, but this time using the diagonal
  $K$-matrix element, $K_{aa}$ from Eq.~(\ref{eq:KMartin}) and
  definition~(\ref{eq:defBranchingRatio})
\begin{equation}
 x^k_{a}=\frac{2}{\Gamma^R_k}\lim_{W\rightarrow M^R_k}\bigg[\left(M^R_k-W\right)\, K_{aa}\bigg],
\end{equation}
  where, as before, all undesired contributions vanish.
\item Steps (iv) and (v) are then repeated for all resonances found in
  (iii).
\end{enumerate}

\section{Results and discussions}
To illustrate the usefulness of our method, resonance parameters from
a unitary, multi-resonance, coupled-channel analysis~\cite{Bat95} have
been extracted. The channels used in the analysis were $\pi N$, $\eta N$,
and an effective two-body channel designated as $\pi^2 N$. Extracted
parameters are given in Table~\ref{tb:resonances}. The proposed model
gives resonance parameters very close to the values obtained by a
complicated method of diagonalizing the matrix of the generalized
Breit-Wigner function denominator, with minimal calculation.
\begin{table}
\begin{tabular}{lccccc}
\hline
\hline
$L_{2I2J}(^{x_{\pi N}/x_{\eta N}/x_{\pi^2 N}}_{M/\Gamma})$ & $M^R$ & $\Gamma^R$ & $x_{\pi N}$ & $x_{\eta N}$ & 
$x_{\pi^2 N}$ \\ 
PDG \protect\cite{PDG04} & [MeV] & [MeV] & [\%] & [\%] & [\%] \\ 
\hline \hline
$S_{11}\left(^{35-55/30-55/1-10}_{1535\pm ^{20}_{15}/150\pm 50}\right)$ & \textbf{1543 } & \textbf{165 }& 
\textbf{39}  
& \textbf{54} & \textbf{7} \\ [-1.8ex]
  & 1553  &  182 &  46 &  50 &  4 \\
$S_{11}\left(^{55-90/3-10/10-20}_{1650\pm ^{30}_{10}/150\pm ^{40}_{5}}\right)$ & \textbf{1680 } & \textbf{233} & 
\textbf{64} & \textbf{16} & \textbf{20 }\\ [-1.8ex]
& 1652  &  202 &  79 &  13 &  8 \\
$S_{11}\left(^{\mathrm{UNKNOWN}}_{\approx 2090/\mathrm{NE}}\right)$ & \textbf{2054} & \textbf{1926} & \textbf{47} & 
\textbf{3} & \textbf{50} \\ [-1.4ex]
& 1812  &  405 &  32 &  22 &  46 \\
\hline
$P_{11}\left(^{60-70/0/30-40}_{1440\pm ^{30}_{10}/350\pm 100}\right)$ & \textbf{1482} & \textbf{541} & \textbf{61} 
& 
\textbf{0} & \textbf{39} \\ [-1.8ex]
& 1439  &  437 &  62 &  0 &  38 \\
$P_{11}\left(^{10-20/6/40-90}_{1710\pm 30 /100\pm ^{150}_{50}}\right)$ & \textbf{1738} & \textbf{170} & \textbf{44}  
& \textbf{12} & \textbf{44 }\\ [-2.ex]
& 1740  &  140 &  28 &  12 &  60 \\
$P_{11}\left(^{\mathrm{UNKNOWN}}_{\approx 2100/\mathrm{NE}}\right)$ & \textbf{2123} & \textbf{379} & \textbf{3} & 
\textbf{83} & \textbf{14 }\\ [-1.4ex]
& 2157  &  355 &  16 &  83 &  1 \\
\hline
$P_{13}\left(^{10-20/0/>70}_{1720\pm ^{30}_{70} /100\pm 50}\right)$ & \textbf{1776} & \textbf{409} & \textbf{20} & 
\textbf{0} & \textbf{80} \\ [-1.8ex]
& 1720  &  244 &  18 &  0 &  82 \\
\hline
$D_{13}\left(^{50-60/0/40-50}_{1520\pm ^{10}_{5}/120\pm ^{15}_{10}}\right)$ & \textbf{1515} & \textbf{121} 
&\textbf{ 
56} &\textbf{ 0} & \textbf{44 }\\ [-1.8ex]
& 1522  &  132 &  55 &  0 &  45 \\
$D_{13}\left(^{50-60/0/40-50}_{1700\pm 50/100\pm 50}\right)$ & \textbf{1818} & \textbf{126} & \textbf{15} & 
\textbf{15 } & \textbf{70} \\ [-1.8ex]
& 1817  &  134 &  9 &  14 &  77 \\
$D_{13}\left(^{\mathrm{UNKNOWN}}_{\approx 2080/\mathrm{NE}}\right)$ & \textbf{2359} & \textbf{1216} & \textbf{26} & 
\textbf{6 } & \textbf{68} \\ [-1.2ex]
& 2048  &  529 &  17 &  8 &  75 \\
\hline
$D_{15}\left(^{40-50/0/50-60}_{1675\pm ^{10}_{5}/150\pm ^{30}_{10}}\right)$ & \textbf{1674} & \textbf{144} & 
\textbf{36} & \textbf{0 } & \textbf{64} \\ [-1.8ex]
& 1679  &  152 &  35 &  0 &  65 \\ 
\hline
$F_{15}\left(^{60-70/0/30-40}_{1680\pm ^{10}_{5}/130\pm 10}\right)$ & \textbf{1682} & \textbf{144} & \textbf{67} & 
\textbf{1} & \textbf{32} \\ [-1.8ex]
& 1680  &  142 &  67 &  0 &  { 33} \\ 
\hline
$F_{17}\left(^{\mathrm{UNKNOWN}}_{\approx 1990/\mathrm{NE}}\right)$ & \textbf{2139 }& \textbf{412} & \textbf{7} & 
\textbf{3} & \textbf{90 }\\ [-1.ex]
& 2262  &  2036 &  3 &  2 &  95 \\
\hline
$G_{17}\left(^{10-20/\mathrm{UNKNOWN}}_{2190\pm ^{10}_{90}/450\pm 100}\right)$ & \textbf{1806 }& \textbf{286 }& 
\textbf{6} & \textbf{0 } & \textbf{94 }\\ [1.8ex]
$G_{17}\left(^{-/-/-}_{-/-}\right)$ & \textbf{2397} & \textbf{1217} & \textbf{16 } & \textbf{0 } & \textbf{84} \\ 
[-1.8ex]
& 2125  &  381 &  18 &  0 & 82 \\
\hline \hline
\end{tabular} 
\caption{Resonance parameters extracted using the $K$-matrix procedure
given in this paper are listed in bold face. The original $T$ matrix
was taken from Ref.~\protect\cite{Bat95} where the channels used were
$\pi N$, $\eta N$, and an effective two-body channel $\pi^2
N$. For comparison, Breit-Wigner resonance parameters from the original
reference are shown below.}
\label{tb:resonances}
\end{table}

\begin{figure*}
\includegraphics[width=4cm]{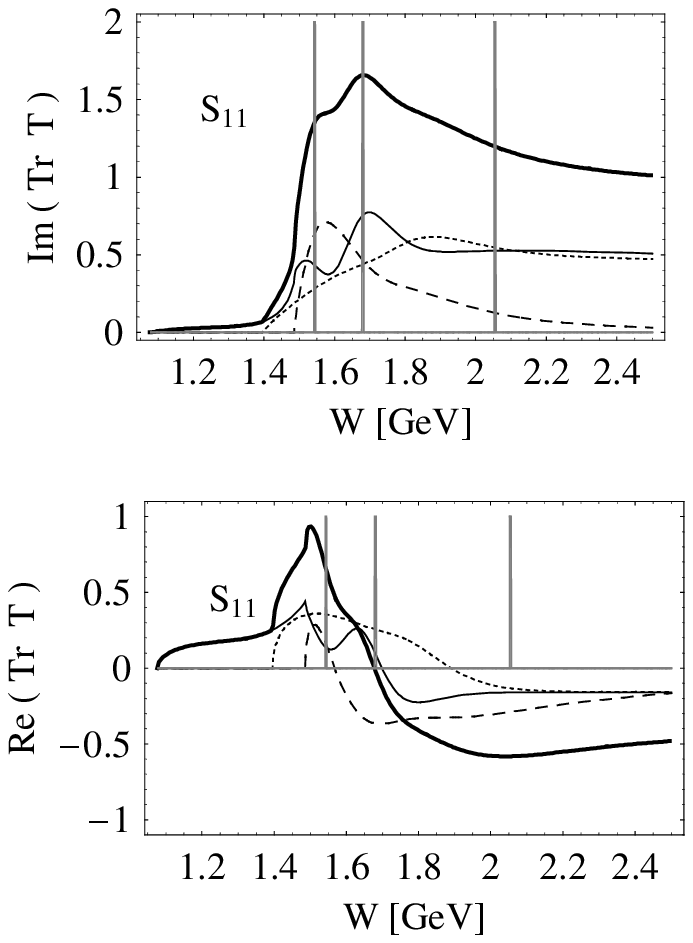}
\includegraphics[width=4cm]{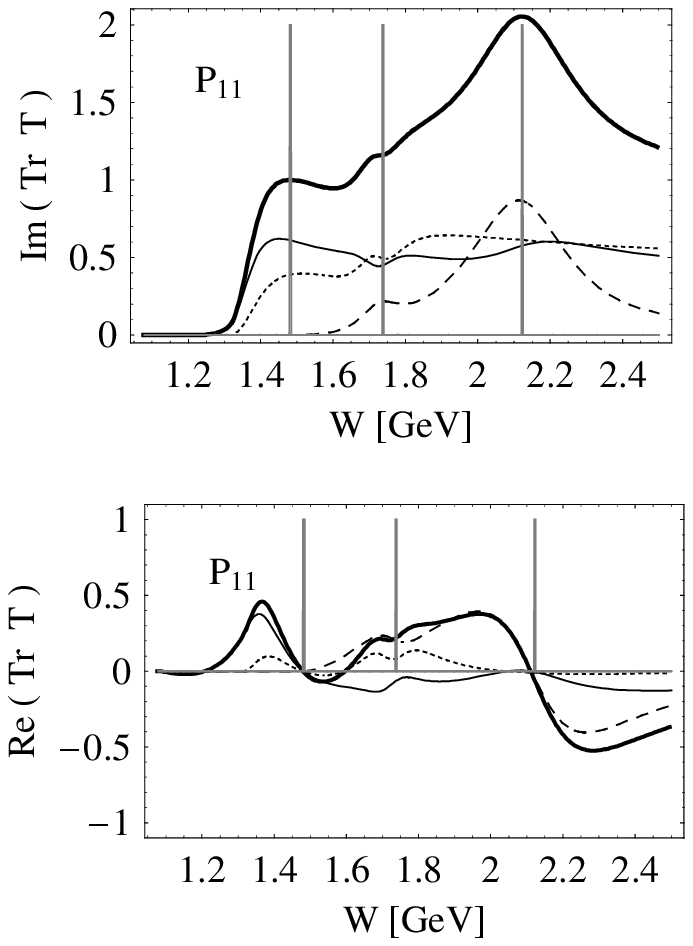}
\includegraphics[width=4cm]{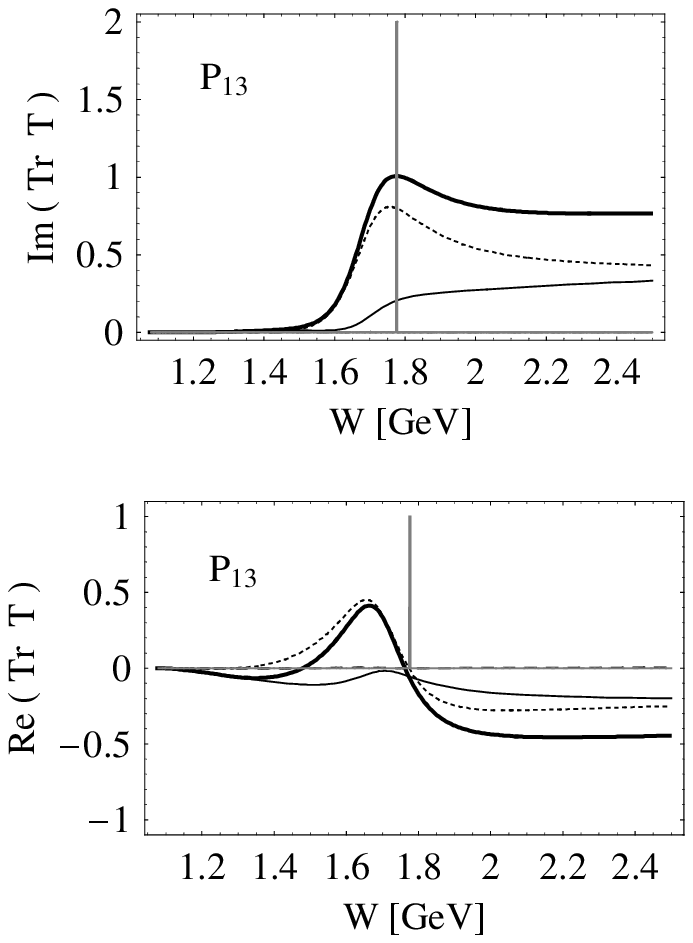}
\includegraphics[width=4cm]{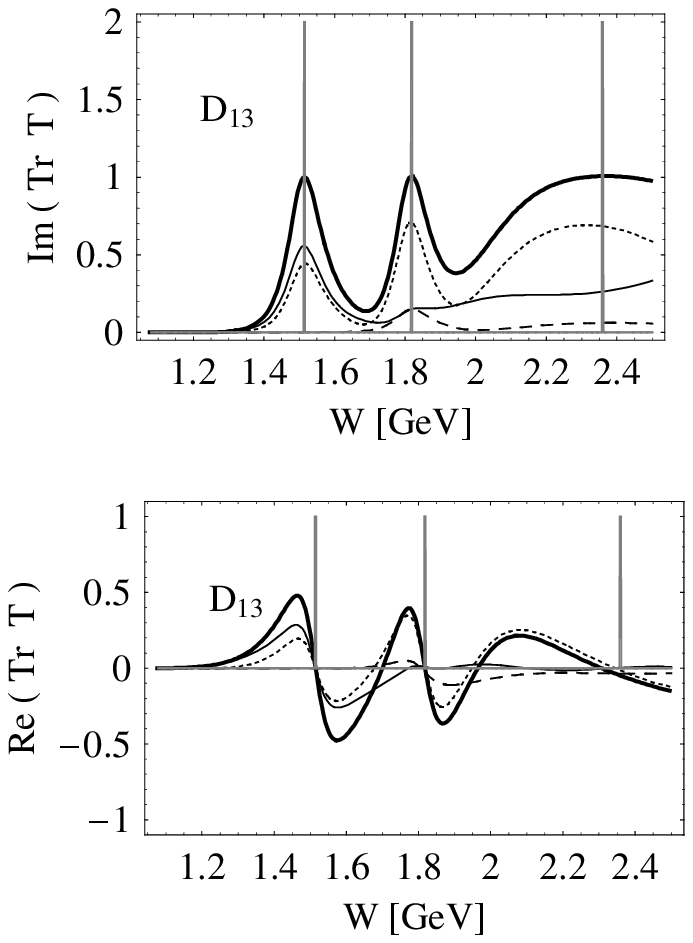}
\\
\includegraphics[width=4cm]{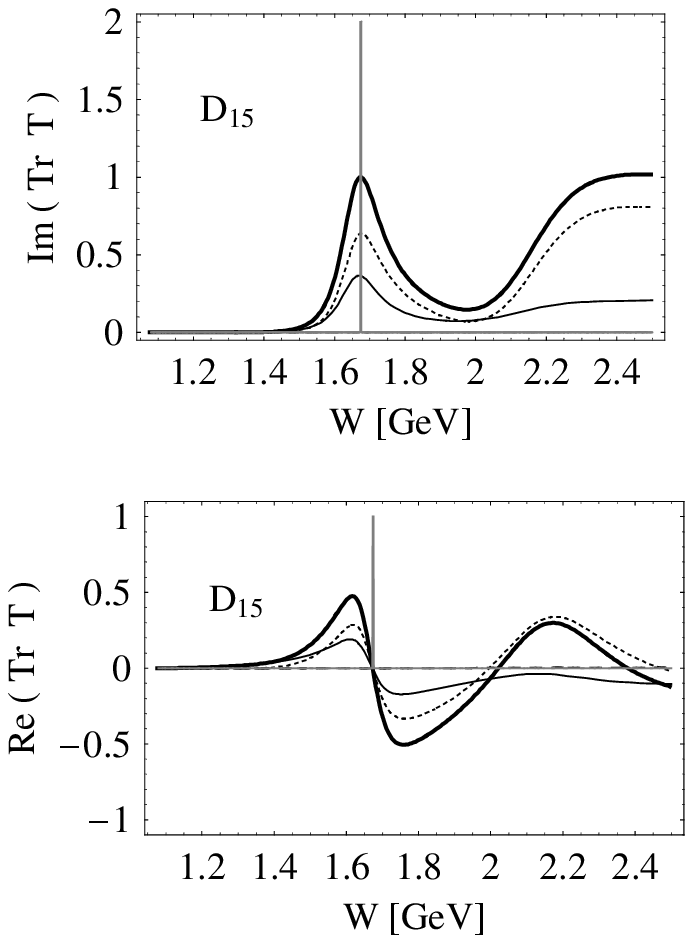}
\includegraphics[width=4cm]{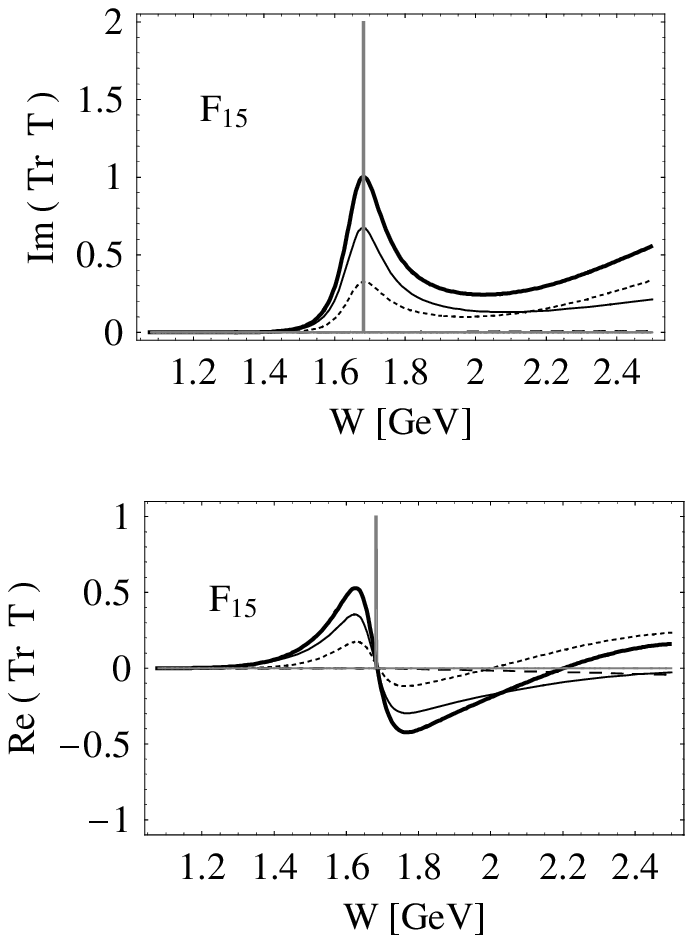}
\includegraphics[width=4cm]{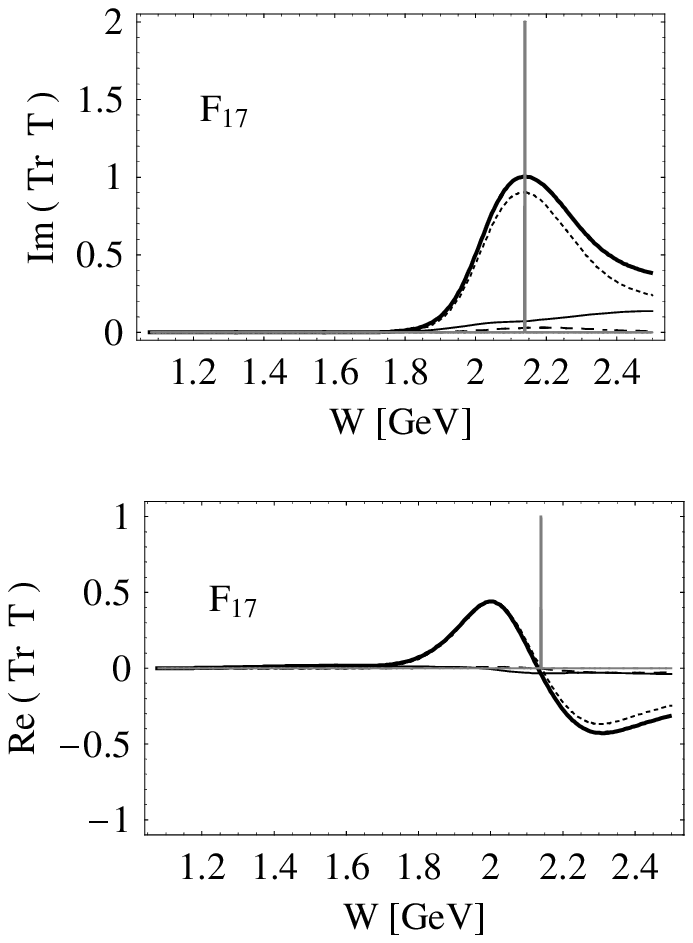}
\includegraphics[width=4cm]{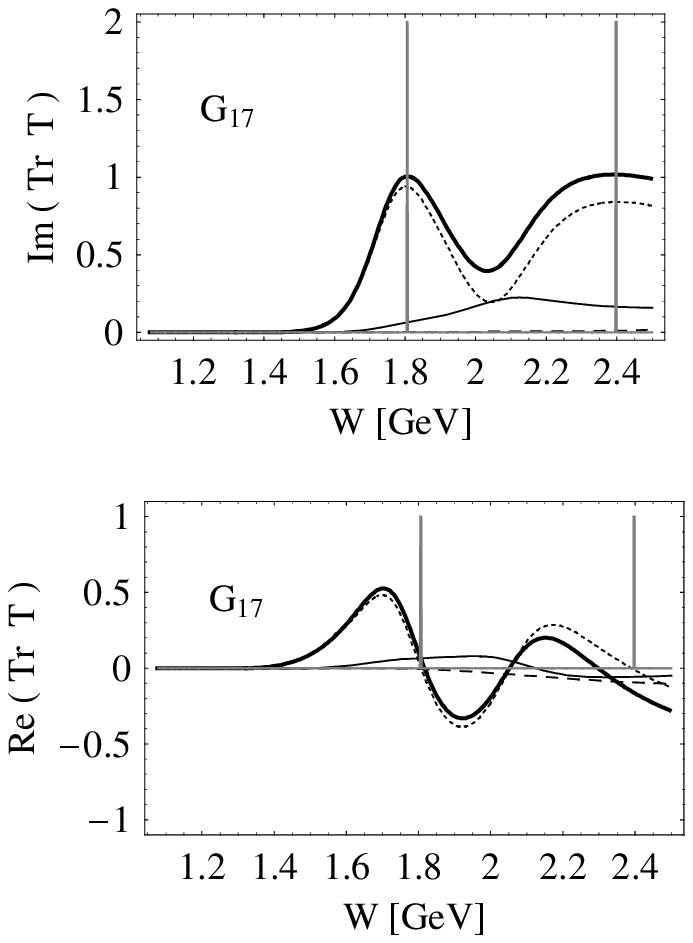}

\caption{The trace of the $T$ matrix and its contributions for partial waves
from $S_{11}$ to $G_{17}$. The thick black line represents the
imaginary part (upper graph) and real part (lower graph) of the trace
of the $T$ matrix. The thin line is the $\pi N$ elastic contribution, the
dashed line shows the contribution from $\eta N$, while the dotted
line gives the effective-channel contribution (unitarity
channel). Gray vertical lines are plotted at the trace of $K$-matrix
pole positions.}
\label{fig:S11-G17}
\end{figure*}

We have also compared the $K$-matrix trace to that of the
$T$ matrix. It can be seen in Fig.~\ref{fig:S11-G17} that the
Breit-Wigner resonance positions obtained by looking for the poles in
$\mathrm{Tr}\,K$ (indicated by gray vertical lines) directly
correspond to the positions of peaks in $\mathrm{Im}(\mathrm{Tr}\, T)$,
and of zeros in $\mathrm{Re}(\mathrm{Tr}\, T)$. The peaks of the
$T$-matrix elements corresponding to individual channels, however,
show a certain deviation from that behavior. This suggests that
fitting individual channels in order to obtain resonance parameters
introduces an uncontrolled error, which is avoided if the trace of the
$T$ matrix is used.

Unexpectedly, and contrary to previous findings, the resonance
parameters produced by the $K$-matrix extraction method presented
here, are in accordance with values obtained by the original analysis
as well as with the $T$-matrix trace. The procedure involves no
fitting, diagonalizing, nor modeling of the energy dependence of the
resonance parameters and background. Furthermore, a model-independent
procedure cannot be given with the $T$-matrix formalism, because
background makes a substantial contribution to the $T$ matrix, even
at an energy equal to the resonance mass, $M^R$. The $T$-matrix background
is removed at a complex energy equal to the $T$-matrix pole
position. This might be the reason why extractions of $T$-matrix poles
work much better than $T$-matrix extractions of Breit-Wigner
parameters. By using the trace of the $K$ matrix, background has been
completely removed from consideration at the resonance energies.

With regard to the differences between the two approaches listed in
Table~\ref{tb:resonances}, it is rather striking that all of them can
be explained by arguments presented in the original analysis. Since an
effective $\pi^2 N$ channel was introduced in~\protect\cite{Bat95} to
parametrize the first inelasticity in each partial wave, the
parameters of low-lying resonances should be much better determined
than those of heavier ones (especially the third resonances in
$S_{11}$ and $D_{13}$). A better quality of parameters is also
expected for resonances that couple more strongly to the measured
channels considered here. Therefore, $N$(1720) $P_{13}$ and the
resonance(s) in $G_{17}$ have unrealistic parameters since they are
completely driven by the effective channel, as can be clearly seen
from Fig.~\ref{fig:S11-G17}. These problems should be removed by the
explicit inclusion of additional channels in the partial-wave
analysis.

The parameters of the two lowest resonances in the $S_{11}$ and
$D_{13}$ partial waves, as well as those of the $D_{15}$, second
$P_{11}$, and $F_{15}$ resonances, are in rough accordance with
quark-model expectations for their masses and partial
widths~\cite{Cap00}, with the exception of the mass of the second
$D_{13}$, which is predicted to be roughly degenerate with the second
$S_{11}$ and the $D_{15}$ resonance. This disagreement could be
explained by the large coupling of this state to the effective
channel. The large width and the somewhat larger mass of the first
$P_{11}$ (Roper) resonance extracted using the $K$-matrix procedure
bring these parameters closer to those of the class of quark-model
calculations based on one-gluon exchange potentials and pair creation
for strong decays.

\section{Conclusions}

We have presented a model-independent method for resonance parameter
extraction using the $K$-matrix formalism. It is shown that real poles
of the $K$ matrix are related to the resonant behavior of the trace of
the $T$ matrix. Our resonance parameter extraction procedure is simple
and straightforward once the full $T$ matrix is known. Unrealistic
extracted parameters for some higher mass resonances point to the need
to include additional channels in partial-wave analyses.

At the energies of the $K$-matrix poles, the influence of background
and channel mixing is eliminated, so only parameter values obtained at
this particular energy should be compared directly to the predictions
of quark model and lattice QCD calculations.

This model-independent procedure cannot be extended to the $T$-matrix
formalism because background makes a substantial contribution to the
$T$ matrix, even at the resonance energies $M^R$. This might be the
reason why methods that extract $T$-matrix poles work much better
than those which extract Breit-Wigner parameters from the
$T$ matrix. By using the trace of the $K$ matrix, the background has
been completely removed from consideration at resonance energies.



\end{document}